\documentclass[aps,pre,amsmath,preprint,showpacs]{revtex4}
\usepackage{graphics,epsfig}
\usepackage{bm}

\begin{document}

\title{Pitchfork and Hopf bifurcation threshold in stochastic equations
with delayed feedback} 
\author{Fran\c{c}oise Lepine and Jorge Vi\~nals}
\affiliation{Department of Physics, McGill University, Montreal, QC H3A 2T8, 
Canada}

\date{\today}

\begin{abstract}
The bifurcation diagram of a model nonlinear Langevin equation with 
delayed feedback is obtained numerically. We observe both direct and
oscillatory bifurcations in different ranges of model parameters.
Below threshold, the stationary distribution function $p(x)$ is a
delta function at the trivial state $x=0$. Above threshold, $p(x) \sim
x^{\alpha}$ at small $x$, with $\alpha = -1$ at threshold, and monotonously
increasing with the value of the control parameter above
threshold. Unlike the case without delayed
feedback, the bifurcation threshold is shifted by fluctuations by an amount
that scales linearly with the noise intensity. With numerical information
about time delayed correlations, we derive an analytic expression for $p(x)$ 
which is in good agreement with the numerical results.
\end{abstract}

\maketitle

We obtain by numerical means the complete bifurcation diagram of a generic 
nonlinear and non Markovian Langevin equation that incorporates the effect of
delayed feedback. Both pitchfork and Hopf bifurcation thresholds are observed
and studied, and the results contrasted with two related limits: The
deterministic limit of a differential delay equation, and the stochastic
bifurcation of the same model without delay.

The study of differential delay equations \cite{re:driver77} is an important 
topic in applied mathematics, with widespread applications in Physics 
(lasers, liquid crystals), control systems in Physiology (neural
and cardiac tissue activity) \cite{re:mackey77,re:glass88}, and Economy
(agricultural commodity prices). Recent interest has arisen in
the mathematical modeling of cellular function at the molecular level,
especially in transcriptional gene regulation \cite{re:hasty01}. Feedback
regulation is a common motif in complex cellular networks,
with delays arising from the complexity of the underlying network, or from the wide
disparity in time scales of the many chemical processes involved in 
regulation \cite{re:glass88}. For example, DNA is transcribed at a rate of 10 to 100 
nucleotides per second, and it may take a delay of the order of minutes 
before the transcription factor appears as a finished product in the cell and
is available for regulation. Significant delays can also be attributable to
the time required for the diffusion of proteins through membranes, so that,
for example, the auto regulated feedback on protein production at time $t$ is
often proportional to protein concentration at time $t-\tau$, where $\tau$ is
known as the delay time. For short delay times, a
reaction may be approximated as being instantaneous, and the system as being
in quasi equilibrium. However, when the delay is comparable to the
characteristic time scales of reaction, the non instantaneous nature of the
interactions can no longer be ignored, and delay terms need to be included in
the governing equations for the network under study
\cite{re:mackey95,re:bratsun05}.

Experimental evidence has been mounting that highlights the importance of
stochastic effects in transcriptional regulation 
\cite{re:kepler01,re:elowitz02,re:bratsun05,re:maheshri07}, not only for
natural networks, but for engineered gene circuits and networks as well
\cite{re:elowitz00,re:hasty00}. 
However, despite the wealth of evidence on the subject, delays in 
stochastic models of metabolic feedback are very often neglected, possibly
because the resulting stochastic equations are no longer Markovian, and hence
rarely tractable analytically. Exceptions include the derivation of a two time Fokker-Planck
equation and the study of its small delay time limit in 
\cite{re:guillouzic99}, and results on the bifurcation of the first
and second moments of a stochastic linear equation with delay
\cite{re:mackey95,re:lei07}. We extend these latter results 
to the analysis of the stationary probability distribution function
of a nonlinear model, and discuss in detail the stability of the solutions
that results from the interplay of delay and stochasticity.

We focus on a canonical form of a nonlinear Langevin equation with
multiplicative or parametric noise and delayed feedback
\begin{equation}
\dot x = ax(t) + bx(t-\tau) - x^3(t) + \xi(t)x(t)
\label{eq:StochmultDDnonlin_ch3}
\end{equation}
where the constant $a$ plays the role of the control parameter, $b$ is
the intensity of a feedback loop of delay $\tau$, and $\xi(t)$ is a white,
Gaussian noise of intensity $D$. The initial condition is a function $\phi(t)$
specified on $t=[-\tau,0]$. We study the stationary probability
distribution function $p(x)$ for a range of values of $a, b$ and $D$.

\begin{figure}
\centering\includegraphics[width=6in]{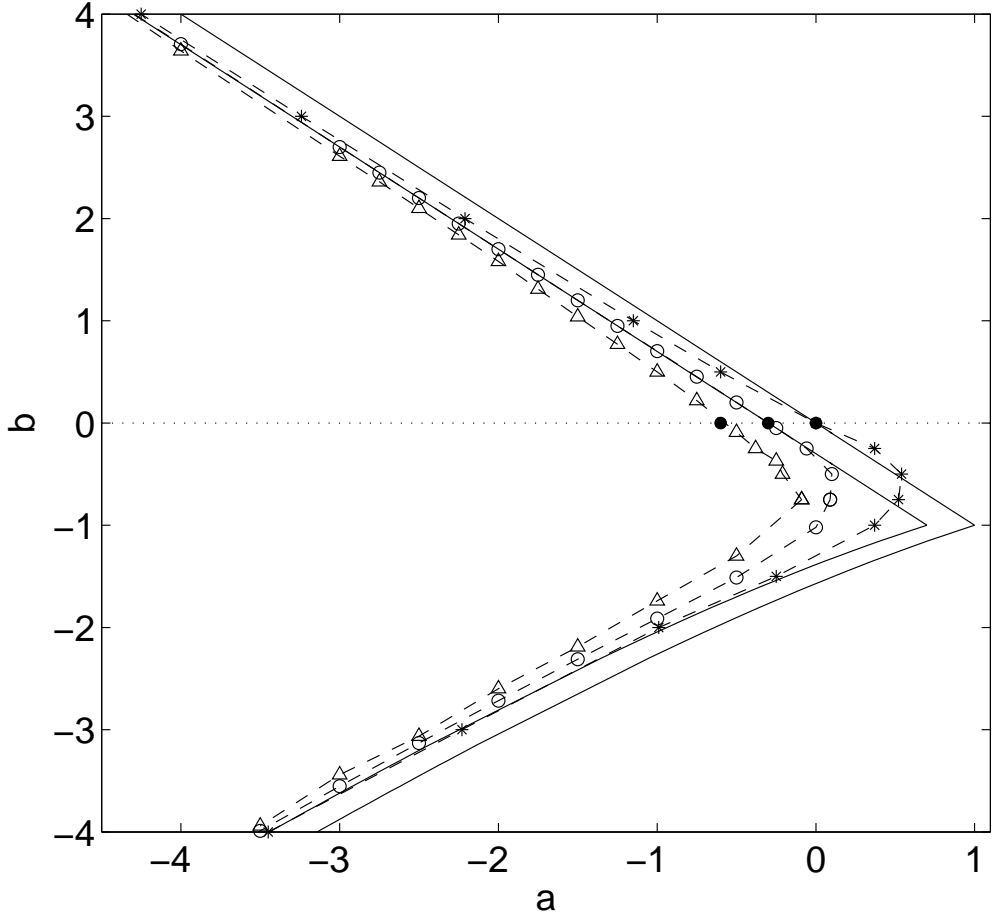}
\caption{Bifurcation diagram of Eq. (\ref{eq:StochmultDDnonlin_ch3}) with
  $\tau = 1$. The outer solid line corresponds to the deterministic limit, the inner solid
  line is the analytic result of ref. \cite{re:mackey95} for $\langle x
  \rangle$, $\circ$ and $\triangle$, are the stability thresholds of $\langle x
  \rangle$  and $\langle x^{2} \rangle$ of the linearized equation
  respectively, and $*$ the threshold obtained from $p(x)$. The three
  points labeled by $\bullet$ on the line $b=0$ are known results for no
  delay. All three are in good agreement with our numerical results.} 
\label{fig:bif_diagram}
\end{figure}

The bifurcation diagram corresponding to Eq. (\ref{eq:StochmultDDnonlin_ch3})
in the absence of noise is known (see, e.g. \cite{re:mackey95}). Linearization
around $x=0$ shows that trajectories decay asymptotically to zero if
\begin{equation}
\textrm{(I)} \,\, b < -a \,\,\textrm{ and, }\,\, \textrm{(II)} \,\, \tau <
  \tau_c = \frac {cos^{-1}(-\frac{a}{b})}{\sqrt{b^2 - a^2}}
\label{eq:det_boundary}
\end{equation}
The boundary separating exponentially decaying solutions from exponentially
growing solutions is shown as the solid line in Fig. (\ref{fig:bif_diagram}). 
The upper branch, (I), is defined by $a_{c} = -b$ and corresponds to a direct
bifurcation (real eigenvalue), whereas the lower branch, (II), corresponds to
a Hopf bifurcation (complex eigenvalue). In both cases, we show in the figure
the case of $\tau = 1$, and hence the lower brach corresponds to $\tau_{c} =
1$ in Eq. (\ref{eq:det_boundary}). The cusp at the intersection of both
boundaries is located at $(a,b) = (1/\tau, -1/\tau)$.



The stochastic bifurcation analysis of Eq. (\ref{eq:StochmultDDnonlin_ch3})
without delay ($b = 0$) is also known
\cite{re:graham82b,re:drolet98,re:sanmiguel99}. Analysis of the linearized
equation leads to the unphysical conclusion that the bifurcation threshold
depends on the order of the statistical moment considered. With the saturating
nonlinearity in Eq. (\ref{eq:StochmultDDnonlin_ch3}), stationary probability
distributions of $x$ can be obtained both below and above threshold, and the
location of the threshold properly determined. The stationary
distribution function of $x$ with $b = 0$ is \cite{re:graham82b}
\begin{eqnarray}
\alpha \leq -1 & & p_{0}(x) = \delta(x) 
\label{eq:delta_function} \\
\alpha > -1 & & p_{0}(x) = N
x^{\alpha}e^{-\frac{x^2}{2D}}
\label{eq:nlin_prob}
\end{eqnarray}
where the exponent $\alpha = a/D - 1$, and $N$ is a normalization constant.
The solution (\ref{eq:nlin_prob}) exists but is not normalizable for $\alpha <
-1$ and hence it is not a physically admissible solution. Therefore
the bifurcation threshold is located at $a = 0$ where $p(x)$ changes from a
delta function at the the origin to a power law at small $x$ with an
exponential cut off at large $x$. In this case, the existence of parametric
fluctuations has no effect on the location of the bifurcation threshold: Both
deterministic and stochastic equations exhibit a bifurcation at $a_{c} = 0$.
In $-1 < \alpha < 0$, $p(x)$ is unimodal
with a divergence at $x=0$, whereas for $\alpha > 0$ the distribution is
bimodal reflecting saturation of $x$.

We now turn to the case of delay, $b \neq 0$.
Analytical results for the stability of the trivial solution $x=0$ of the {\em
linearization} of Eq. (\ref{eq:StochmultDDnonlin_ch3}) have been given in
\cite{re:mackey95,re:lei07}, and are shown in Fig. (\ref{fig:bif_diagram}). 
The bifurcation threshold of the first moment is shifted relative to the
deterministic limit, only bounds have been given for the second
moment \cite{re:lei07}, and no results are available for $p(x)$. Given the
anomalous behavior described above for the stochastic bifurcation of the
linear equation with $b = 0$, it is of interest to determine $p(x)$ for the
full model of Eq. (\ref{eq:StochmultDDnonlin_ch3}). Unfortunately, the non
Markovian character of this equation has precluded progress along these lines
\cite{re:guillouzic99}. 

We have first extended an existing high order algorithm for the integration of
stochastic differential equations \cite{re:fox91} to the case of delay. The algorithm
needs to take into account trajectories into the past for an interval $\tau$,
and also new contributions from the stochastic terms that result from the
coupling to the delayed feedback. In the numerical results shown below we
employ, for technical reasons, an Ornstein-Uhlenbeck stochastic process $\xi(t)$
with intensity $D$ and correlation time $\Delta t$, the same as the time
step in the discretization of Eq. (\ref{eq:StochmultDDnonlin_ch3}). The
initial condition considered in all our calculations is a white and Gaussian
random process in $(-\tau,0)$ of zero mean and intensity 1. The
time step used is the numerical integration is $\Delta t = 0.01$.

\begin{figure}[h]
\centering\includegraphics[width=6in]{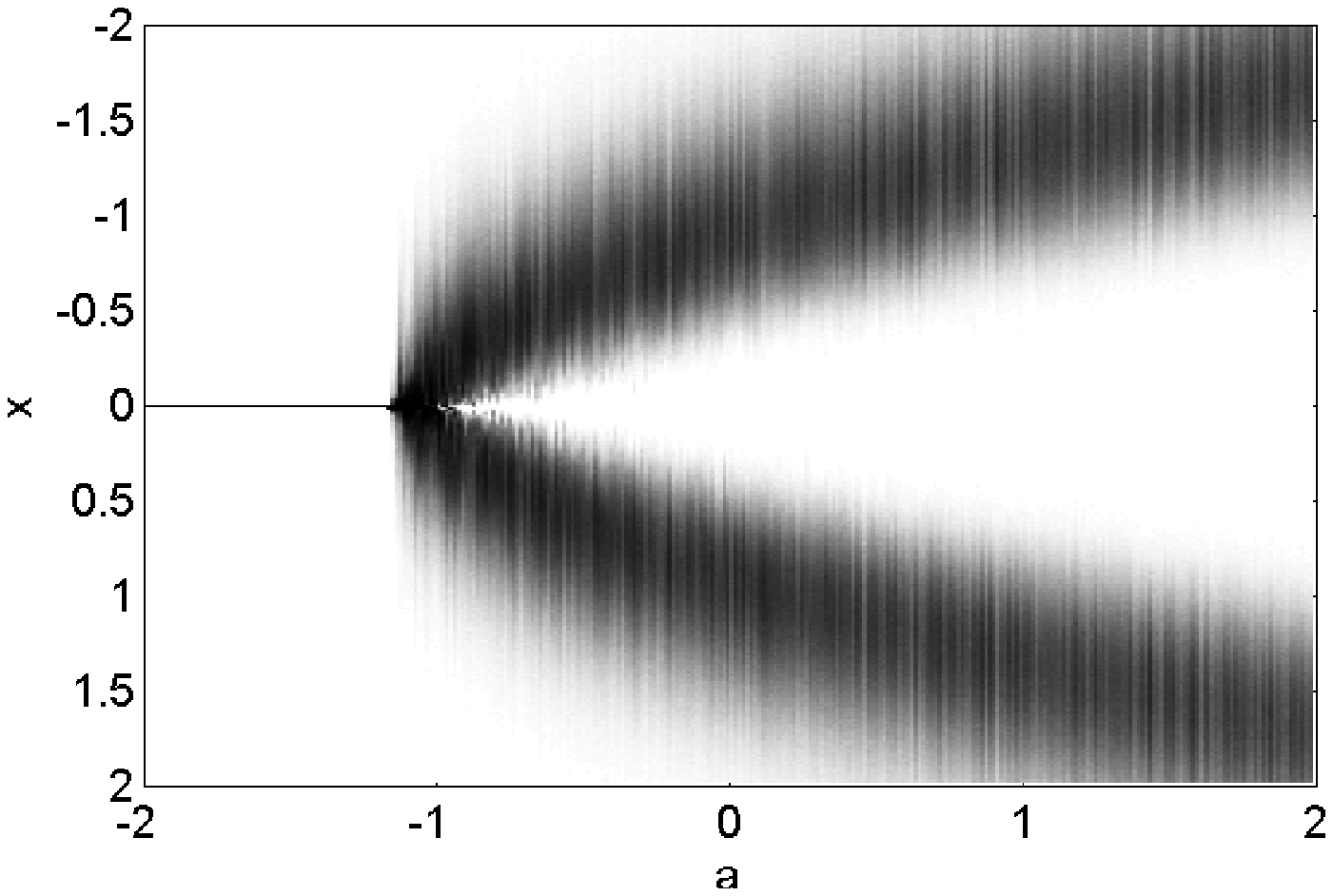}
\caption{Long time histogram of $x$ (in grey scale) as a function of the control
parameter $a$ from Eq. (\ref{eq:StochmultDDnonlin_ch3}) with $b = 1$, $\tau =
1$, and $D = 0.3$.  The histograms have been collected for $t \in (50, 80)$
and further averaged over 60 independent runs.}
\label{fig:histogram}
\end{figure}

\begin{figure}[h]
\centering\includegraphics[width=4in]{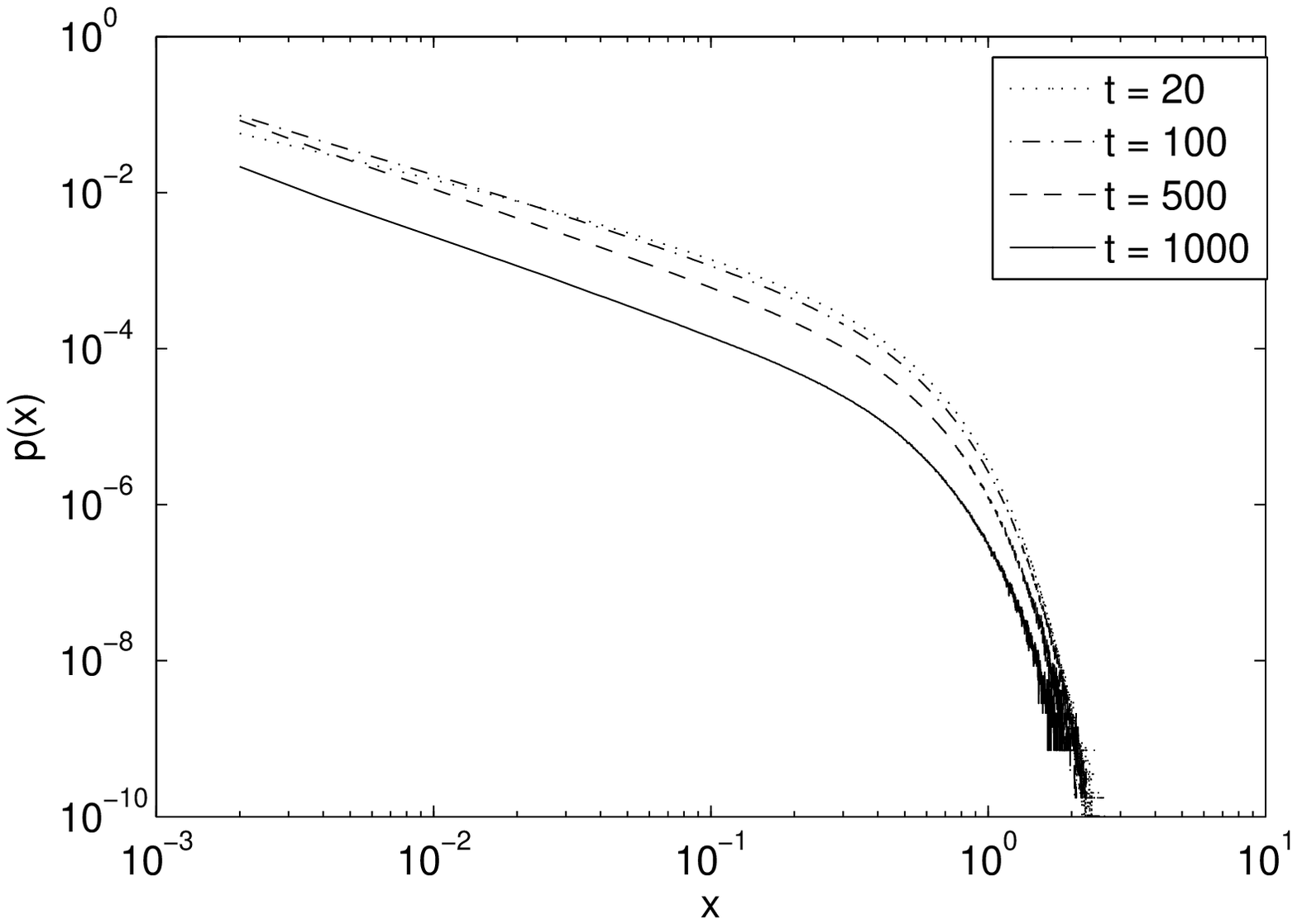} 
\centering\includegraphics[width=4in]{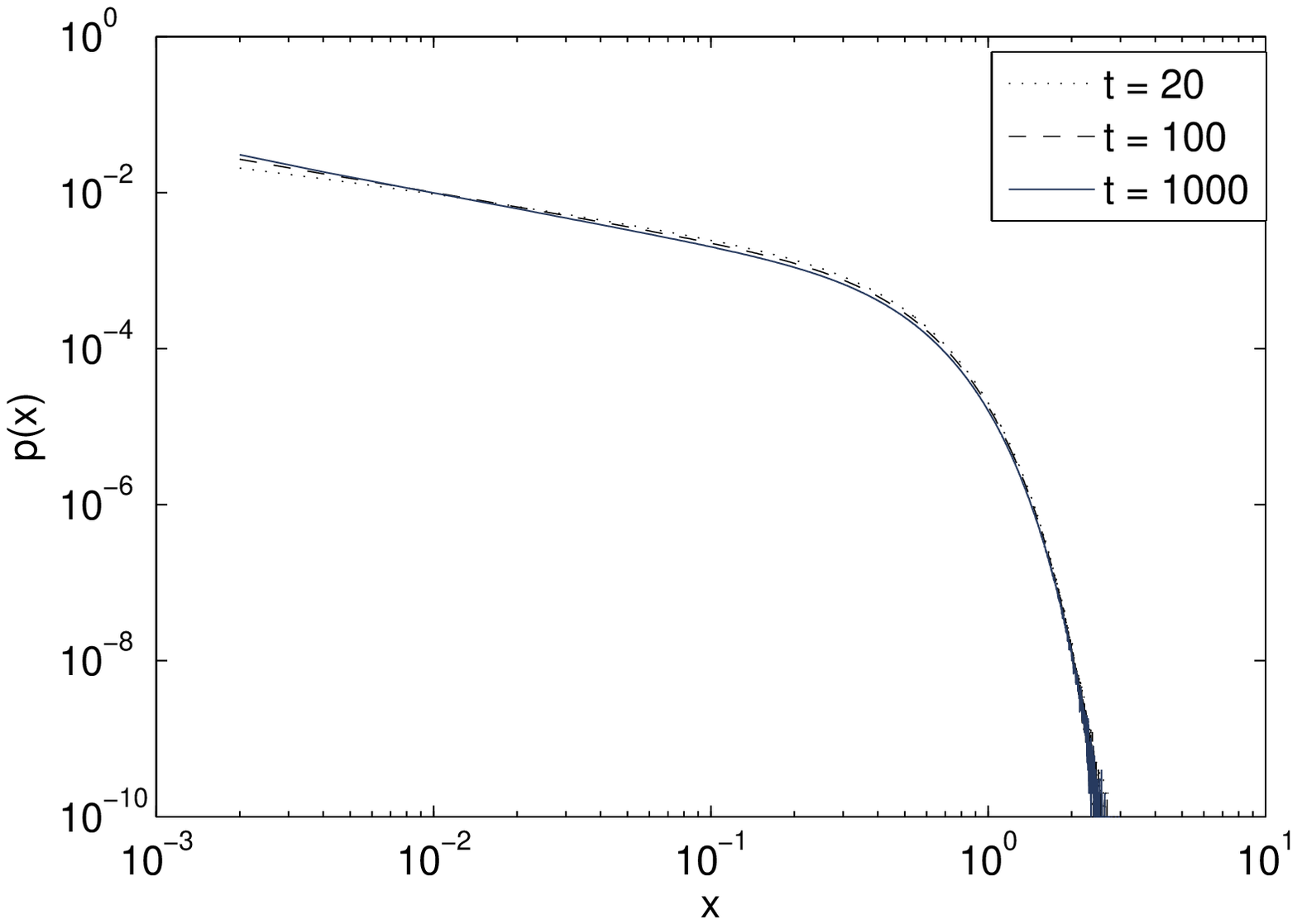}
\centering\includegraphics[width=4in]{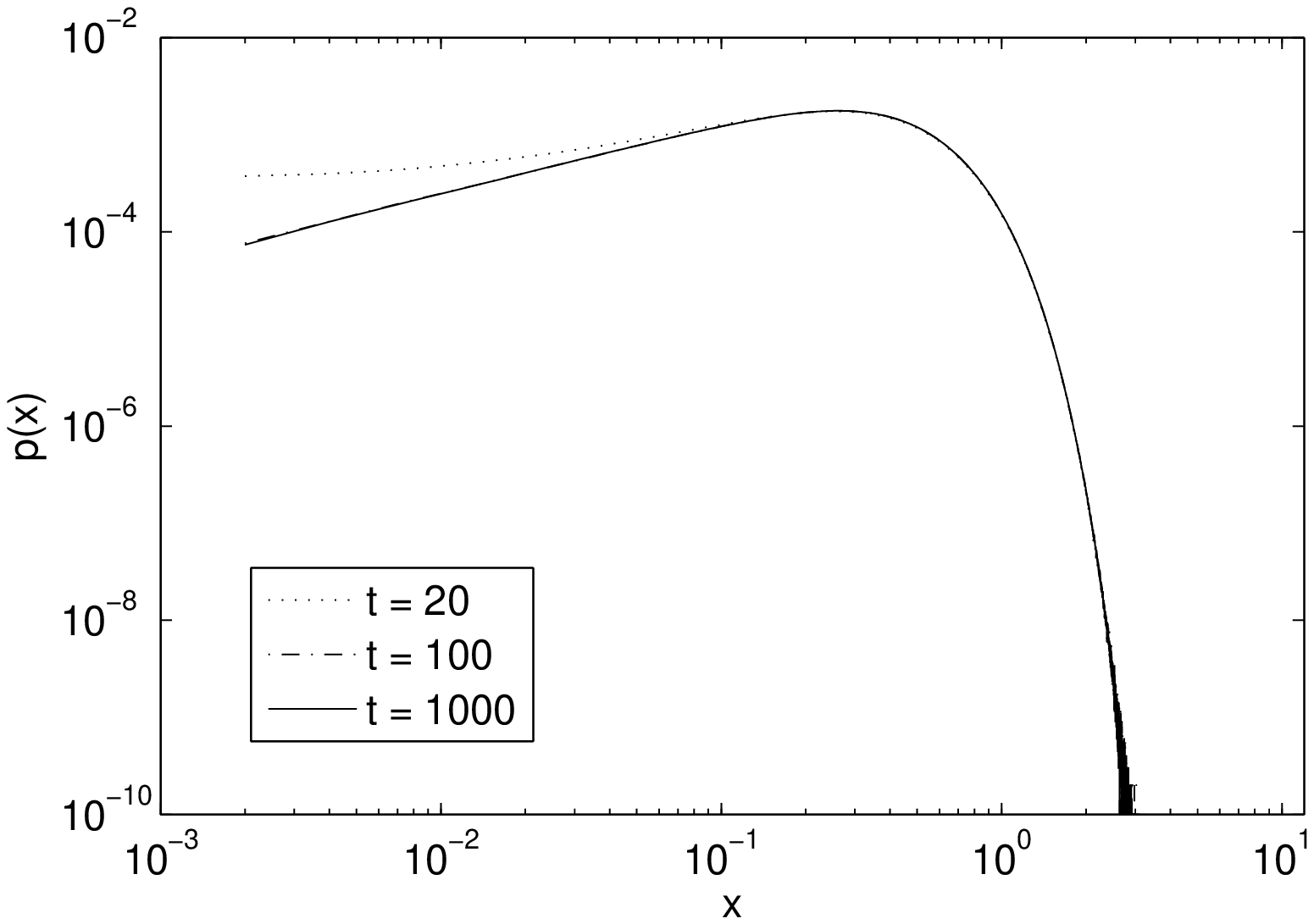} 
\caption{Probability distribution function $p(x)$ for $b = 1$, 
$\tau = 1$ and $D = 0.3$ computed in the time interval $t \in (50,80)$ and
  averaged over $10^{6}$ independent realizations. The values of the
  control parameter shown are: (top) $a = -1.2$, $\alpha \simeq -1.25$, 
(center) $a = -1.1$, $\alpha \simeq -0.73$, and (bottom) $a = -0.9$, $\alpha
  \simeq 0.72$.}
\label{fig:stationary_distribution}
\end{figure}

\begin{figure}[h]
\centering\includegraphics[width=6in]{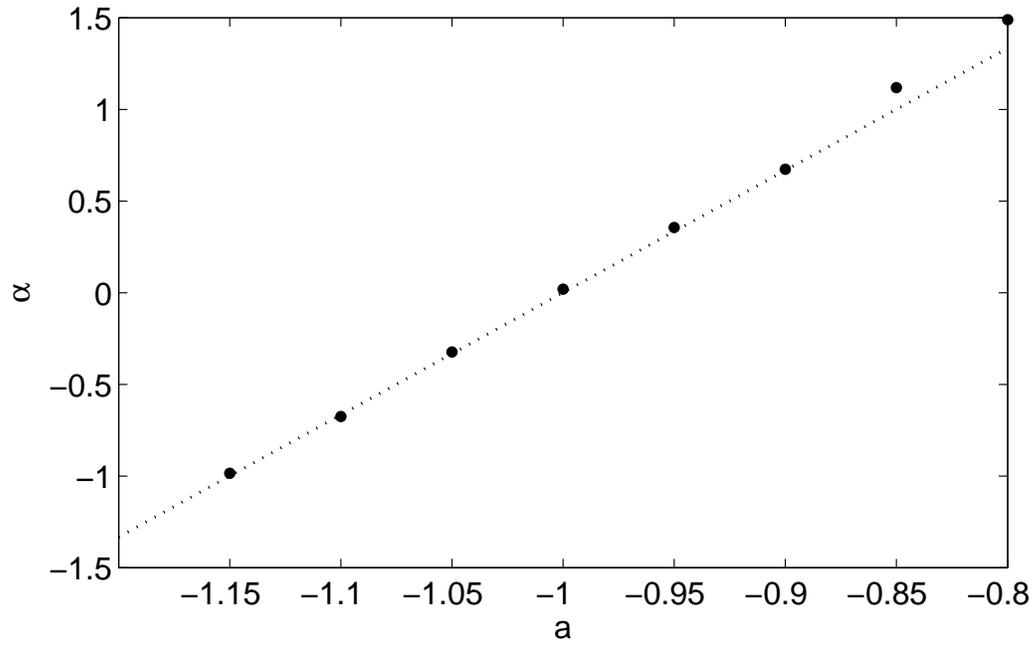}
\caption{Fitted value of $\alpha$ as a function of $a$. We define the 
bifurcation threshold when $\alpha = -1$, or $a_{c} \simeq -1.15$ for this
parameter set ($b = 1$, $\tau = 1$, and $D = 0.3$). The line is obtained from
our empirical determination of the Fokker-Planck equation. There are no
adjustable parameters.}
\label{fig:alpha_a}
\end{figure}

A qualitative view of the bifurcation is given in Fig. \ref{fig:histogram},
which shows a histogram of $x$ once trajectories have reached a statistical
steady state. For $a \lesssim -1$, the histogram is approximately a delta
function at $x=0$. At a critical value $a_{c} \approx -1$, the
bifurcation point, a broad distribution emerges, although the most likely
value remains $x = 0$. At larger values of $a$, the histogram becomes
bimodal. This histogram shown corresponds to the direct bifurcation branch, but
a similar graph is obtained for the Hopf bifurcation. Our results for the
distribution function $p(x)$ in these three ranges of values of $a$
are shown in Fig. \ref{fig:stationary_distribution}. For $a < a_{c}$ $p(x)$ is
approximately a power law with effective exponent $\alpha < -1$, but with a
growing amplitude of $p(x)$ at $x \approx  0$ (not shown in the
figure). Because of normalization, this growth implies a decaying amplitude for
finite $x$, signaling a long transient leading to the build up of the delta
function at $x=0$. Interestingly, the effective power law in the figure
$\alpha \approx -1.2 < -1$, indicating that $p(x)$ would not be
normalizable. For $a > a_{c}$ we do obtain a time independent distribution
with $-1 < \alpha < 0$. This distribution is normalizable, and represents the
stationary distribution above threshold. We finally show
$p(x)$ in the range of $a$ for which the distribution is bimodal. The
probability of the most likely value is approximately constant, but we still
observe some transients in the vicinity of $x=0$. Figure \ref{fig:alpha_a} shows
the results of a power law fit to $p(x)$ as a function of the control
parameter $a$. We observe a smooth variation of the exponent $\alpha$ with $a$
that allows a convenient determination of $a_{c}$, the value of $a$ for which
$\alpha = -1$. This is the method that we have used to determine the
bifurcation threshold in all the results presented below.

\begin{figure}[h]
\centering\includegraphics[width=6in]{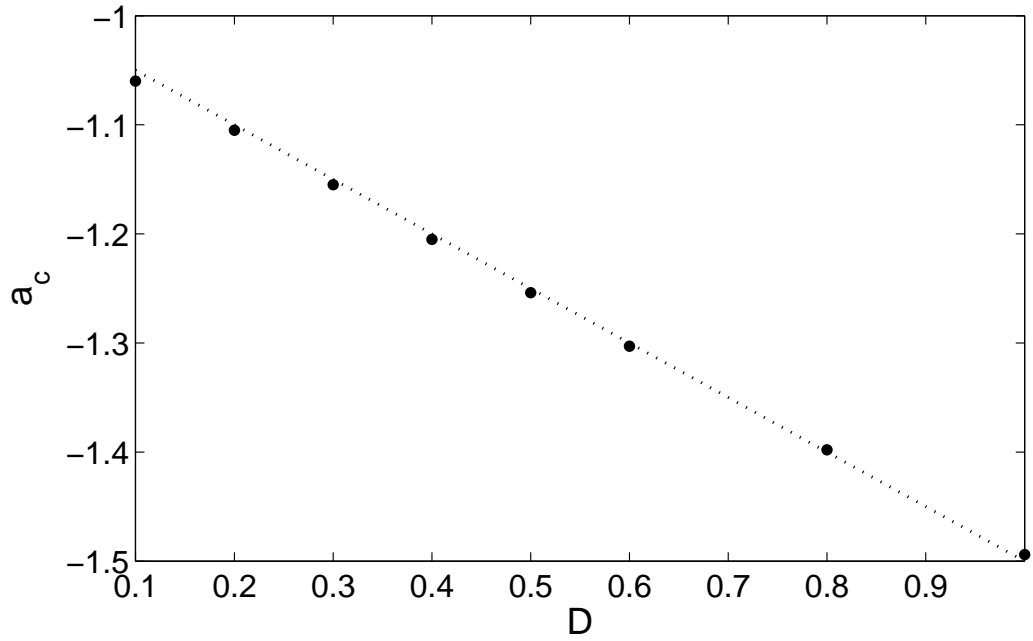}
\caption{Bifurcation threshold $a_{c}$ from $p(x)$ as a function of noise
  intensity $D$. The parameters used are $b = 1, \tau = 1$, time averages for
  $t \in (300,350)$ and over 150,000 independent realizations.The line is obtained from
our empirical determination of the Fokker-Planck equation. There are no
adjustable parameters.}
\label{fig:noise_shift}
\end{figure}

We summarize our results for the bifurcation diagram of
Eq. (\ref{eq:StochmultDDnonlin_ch3})in Fig. \ref{fig:bif_diagram}.
The analytic results for the threshold without noise ($\xi = 0$) are shown for
reference, as well as the threshold of $\langle x \rangle$ of the {\em
linearized} equation \cite{re:mackey95}, and our numerical estimate. Except in
the vicinity of the multi critical point $(a,b) = (1/\tau,
-1/\tau)$ our results are in excellent agreement with the analytic
calculation of the linear equation, thus validating the accuracy of the
numerical algorithm. As one approaches the point $(1/\tau, -1/\tau)$ the Hopf
frequency approaches zero and it is necessary to integrate the differential
equation up to very long times
to differentiate between an unstable trivial solution or an oscillation with a
very long period. Since bounds on the threshold from $\langle x^{2} \rangle$
for the linearized equation have been given in the literature \cite{re:lei07}, we
also show in Fig. \ref{fig:bif_diagram} the threshold for this moment obtained
numerically. Our numerical results do agree with the known
threshold for the special point of no delay $b =
0$ given in \cite{re:graham82b}. Interestingly, the thresholds for
$\langle x \rangle$ and $\langle x^{2} \rangle$ converge to the same values
when one moves away from the point $(1/\tau, -1/\tau)$, both in the direct and
Hopf bifurcation branches. The figure also
presents our results for bifurcation threshold defined directly
from the stationary probability distribution function as discussed above. 
Our conclusion is that the stochastic threshold is shifted away from the deterministic threshold
except in the special point of no delay ($b = 0$), both for the direct and Hopf
bifurcations. This threshold also agrees with that of the first and second
moments of the linearized equation in the range of parameters in which both agree.
Figure \ref{fig:noise_shift} shows the dependence of the shift
in threshold as a function of the noise intensity $D$. In analogy
with the case of no delay, we find a linear dependence in $D$.

We next turn to the equation for $p(x)$ that follows from
Eq. (\ref{eq:StochmultDDnonlin_ch3}). The difficulty in obtaining a closed,
analytic expression lies in the need to find the joint probability distribution
$p(x(t),x(t-\tau))$. When $\tau$ is larger than the correlation time of $x$,
one can assume statistical independence between $x(t)$ and $x(t-\tau)$, or
$p(x(t),x(t-\tau))  = p(x(t))p(x(t-\tau))$, an approximation that has been
considered in the literature (e.g., ref. \cite{re:bratsun05}). However, this
assumption does not hold near a bifurcation since characteristic correlation
times diverge. Instead we write $p(x(t),x(t-\tau)) =
p(x(t-\tau)|x(t))p(x(t))$, where $p(x(t-\tau)|x(t))$ is the conditional
probability of finding $x(t-\tau)$ at $t-\tau$ given $x(t)$ at time $t$. We
have then derived a the Fokker-Planck equation for $p(x)$ that requires only the
determination of $\langle x(t-\tau)|x(t) \rangle$, a correlation which we have not been able
to compute analytically.  We can, however, examine its behavior numerically. We
have found that it reaches a stationary function (independent of $t$), and
that for small $x(t)$ is well described by $\langle x_(t-\tau)|x(t) \rangle =
(1+a+b+D) x(t)$ (for $\tau = 1$). With this empirical relation and some
straightforward algebra, we recover the same solution for $p(x)$ given in
Eq. (\ref{eq:nlin_prob}) but with $\alpha= (a+b(1+a+b+D)-D)/D$.
This solution closely agrees with our numerical results of $\alpha$ versus $a$
as shown by the solid line in Fig. \ref{fig:alpha_a},
and with the value of the shift in $a_{c}$ as a function of the noise intensity
$D$ shown in Fig. \ref{fig:noise_shift}.

In summary, many of the qualitative features of stochastic bifurcations under
multiplicative noise are preserved under the addition of a delayed
feedback. First, the bifurcation remains sharp. Since it is commonly assumed in the
literature that the correlation time of $x$ $\tau_{x} \ll \tau$, the delay
term in Eq. (\ref{eq:StochmultDDnonlin_ch3})) would effectively act as an
additive source of noise, leading perhaps to an imperfect bifurcation. We have
shown this not to be the case. Second, and in agreement with the case of no
delay ($b=0$), we observe that the moments of the linearized equation
bifurcate at different values of the control parameter, which are themselves
different from the threshold predicted from the distribution function of the
full nonlinear equation. In contrast with the case of $b = 0$, however, the
existence of delay introduces a shift in the bifurcation threshold, both for 
direct and Hopf bifurcations, shift that goes to zero as $b \rightarrow
0$. The magnitude of the shift scales linearly with the noise
intensity $D$. Finally, we have empirically derived the stationary solution
of the distribution $p(x)$ that agrees with our numerical determination of the
bifurcation threshold. In view of our results, care must be exercised when
analyzing bifurcation thresholds in numerical simulations of model gene
regulatory networks when the analysis is based on the calculation of moments. 

This research has been supported by NSERC Canada.

\bibliographystyle{prsty}
\bibliography{/home/vinals/mss/references}

\newpage

\newpage

\newpage

\newpage

\newpage

\end{document}